\documentclass[aps,prb,twocolumn]{revtex4-1}
\usepackage{color}
\usepackage{graphicx}
\usepackage{cancel}
\usepackage{amsfonts}
\usepackage{amssymb}
\usepackage{amsthm}
\usepackage{ulem}
\usepackage{url}
\usepackage{hyperref}
\usepackage{float}
\usepackage{multirow}  
\usepackage{dcolumn}   
\usepackage{bm}        
\usepackage{enumerate}
\usepackage{mathtools}

\usepackage[dvipsnames]{xcolor}
\hypersetup{
	colorlinks,
	linkcolor={green!50!black},
	urlcolor={blue!80!black},
	citecolor   = {red!50!black} 
}

\def \bver   {\mbox{\boldmath$\it{b}$}}

\def\z{{\bf z}}

\def  \bsig    {\mbox{\boldmath$\sigma$}}

\usepackage{natbib}
\begin{document}

\title{Bilinear magnetoresistance in topological insulators: role of magnetic disorder}
\author{A. N. Zarezad}
 \author{A. Dyrdał$^{\ast}$}
\address{Department of Mesoscopic Physics, ISQI, Faculty of Physics, Adam Mickiewicz University,\\ ul. Uniwersytetu Poznańskiego 2, 61-614 Pozna\'n, Poland}
\email[]{ e-mail: adyrdal@amu.edu.pl}

\begin{abstract}
Bilinear magnetoresistance is a nonlinear transport phenomenon that scales linearly with the electric and magnetic fields, and appears in nonmagnetic systems with strong spin-orbit coupling, such as topological insulators (TIs). Using the semiclassical Boltzmann theory and generalized relaxation time approximation, we consider  in detail the bilinear magnetoresistance in an effective model describing surface states of three-dimensional topological insulators.
We show  that the presence of magnetic impurities remarkably modifies the BMR signal. In general, scattering on magnetic impurities reduces magnitude of BMR. Apart from this, an additional modulation of the angular dependence of BMR appears
when the spin-dependent component of the impurity potential dominates the scalar one.
\end{abstract}

\date{\today}
\maketitle

\section{Introduction}
Strong spin-orbit interaction in  nonmagnetic systems leads to a variety of transport phenomena, such as the spin Hall effect (SHE)~\cite{Hirsch,Sinova2012,Sinova2015}  or current-induced spin polarization (CISP)~\cite{Edelstein,Dyakonov} (a.k. Edelstein effect), that enable pure electrical manipulation of the spin degree of freedom.  Both SHE and CISP are the main mechanisms of efficient spin-to-charge interconversion \cite{Soumyanarayanan2016Nov,Han2018}. They also give rise, among others, to spin torques (i.e., spin Hall torque and spin-orbit torque) and magnetization switching, domain wall manipulation, ferromagnetic resonance, as well as to various magnetoresistance effects \cite{Manchon_RMP2019,Garello2013,Avci2018,Chen_2013,Lv2018}. For example, in the case of spin Hall magnetoresistance (SMR) \cite{Avci2018,Chen_2013}, the longitudinal electric resistance is affected by the spin Hall effect and its inverse counterpart. SMR has been observed in layered structures consisting of a conducting nonmagnetic (NM) layer adjacent to a thin ferromagnetic (FM) film. A nonequilibrium spin accumulation (due to SHE) appears then at the interface, and this spin accumulation  may interact with the magnetic layer via the spin-transfer torque.
Effectively, a part of spin current diffuses across the interface into the ferromagnetic layer, while another part of the spin current is reflected at the interface back to the nonmagnetic layer and becomes converted into a charge current due to the inverse SHE. The back-converted charge current contributes to the longitudinal current and affects the electric resistance, which becomes dependent on the magnetization orientation in the ferromagnetic layer. A similar magnetization dependent magnetoresistance is also expected due to CISP and spin-orbit torque at the interface \cite{Lv2018}.

Interestingly, in systems with FM/NM interfaces, where NM layer contains spin-orbit interaction, one can observe the so-called unidirectional magnetoresistance (UMR)~\cite{Avci2018,Lv2018}. UMR is a nonlinear phenomenon and appears when the resistance contains a term that scales linearly with charge current density and magnetization. This phenomenon is promising for a  new generation of magnetoresistive sensors and logic elements.

Recently a new type of nonlinear magnetoresistance has been identified in a single nonmagnetic layer, that is  called bilinear magnetoresistance (BMR). BMR scales linearly with the external electric and magnetic fields, and has been observed in nonmagnetic materials with strong spin-orbit interaction, such as topological insulators, thin films of germanium, and 2DEG at the interface of perovskite oxides \cite{Guillet,He2018,Vaz_PRMat2020}.
Two main mechanisms of BMR have been discussed as far. The first mechanism has been recognized in $\mathrm{Bi_{2}Se_{3}}$, a 3D topological insulator with strong hexagonal warping \cite{He2018}. According to the theory provided by Zhang and Vignale \cite{He2018,ZhangVignale2018}, BMR in such TIs appears due to nonlinear (quadratic in the electric field) spin currents related to the second-order correction to the distribution function. In the presence of an external magnetic field, these spin currents are partially converted to the charge current that contributes to the unidirectional magnetoresistance signal. However, this second-order charge current disappears in the absence of hexagonal warping. On the other side, it is known that BMR can also appear in systems with isotropic Fermi contours. In such a case, BMR can be a consequence of nonequilibrium spin polarization and its interplay with certain relaxation mechanisms, such as those related to the structural defects and spin-momentum locking inhomogeneity \cite{Dyrdal_PRL2020}.

In this paper we consider BMR in 3D topological insulators with isotropic Fermi contours. We focus on the influence of  magnetic impurities on BMR. It is well known that magnetic impurities are responsible for anisotropic magnetoresistance in systems with strong spin-orbit interaction. Thus, it is important to understand how anisotropic scattering on magnetic impurities affects the BMR signal. This is an interesting issue, as the magnetic disorder or adatoms are often observed on surfaces of TIs. We show that scattering on magnetic impurities generally reduces magnitude of the BMR generated for instance by scattering on spin-momentum locking inhomogeneities. Moreover, such scattering  introduces additional modulation of BMR with the angle between current and magnetic field orientations. 
Accordingly, we assume that BMR is due to spin-momentum locking inhomogeneity and nonequilibrium spin polarization, as discussed previously by Dyrdal et al. \cite{Dyrdal_PRL2020}. The manuscript is organized as follows. In Sec. 2 we introduce the model and scattering mechanisms present in the system. In Sec. 3 the generalized Boltzmann equation and its solution for the model presented in Sec.2 are described in detail. Finally, in Sec. 4 we present and discuss numerical results on the
asymmetric
components of magnetoresistance, i.e.,
on the bilinear magnetoresistance (BMR). Summary and final conclusions are presented in Sec. 5.

\section{Model}

We consider the minimal model (i.e., we neglect the terms describing particle-hole asymmetry and hexagonal warping) describing surface states in three-dimensional topological insulator in the presence of external in-plane magnetic field and disorder. Accordingly, we write the relevant Hamiltonian  as
\begin{equation}
\label{eq:H}
\hat{H}=\hat{H_{0}}+\hat{V}^{\mathrm{im}},
\end{equation}
where the unperturbed Hamiltonian takes the following form:
\begin{equation}
\label{eq:H0}
\hat{H_{0}} = \upsilon (\mathbf{k} \times\hat{\z})\cdot\bsig + (\mathbf{B} + \mathcal{J} \mathbf{S}^{E}) \cdot \bsig .
\end{equation}
The first term in $\hat{H}_{0}$ describes the surface states of TI with $\upsilon =\hbar  v_F$ ($v_F$ being the Fermi velocity)  and $\mathbf{k} = (k_{x} , k_{y})$ denoting the wavevector \cite{Adam2012}. The second component describes an effective magnetic field $\mathbf{B}_{\mathrm{eff}} = \mathbf{B} + \mathcal{J} \mathbf{S}^{(E)}$ acting of the surface electrons. Thus, $\mathbf{B} \cdot \bsig$ is the Zeeman term describing the influence of external in-plane magnetic field, $\mathbf{B}= (B_x,B_y,0) = B (\cos\phi_{B}, \sin \phi_{B}, 0)$ (note that $\mathbf{B}$ is in energy units, i.e. $B = g \mu_{B} \mathcal{B}$, and  $\bsig=(\sigma_x,\sigma_y,\sigma_z)$ is vector of Pauli matrices for surface electrons). In turn,  the term $\mathcal{J} \mathbf{S}^{(E)}\cdot \bsig$ describes
surface electrons
in an effective spin-orbit field,  $\mathcal{J} \mathbf{S}^{(E)}$, which is proportional to the non-equilibrium spin polarization $\mathbf{S}^{(E)}$, with $\mathcal{J} = - 8\pi v_{F}/k_{F}$ playing the role of coupling parameter.
Note that the non-equilibrium spin polarization considered here is driven by an external electric field in the presence of spin-orbit coupling and is also known as inverse spin-galvanic effect or Edelstein effect.  Without loosing generality,  we assume in this work that the external electric field is applied along the $x$-axis,  and therefore the nonequilibrium spin polarization has only y-component, i.e., $\mathcal{J} \mathbf{S}^{(E)}\cdot \bsig \rightarrow \mathcal{J} S^{(E)}_y  \sigma_{y}$.

The second term in~Eq.(\ref{eq:H}), $\hat{V}^{\mathrm{im}}$, describes possible scattering processes in the system.
Since the surfaces of TIs are in general not perfect and possess structural defects that inherently include spin-orbit interaction, we consider the scattering potential describing this type of inhomogeneity \cite{Dyrdal_PRL2020,Sherman2003,Golub2004}:
\begin{equation}
 \hat{V}^{\mathrm{SOC}}(\mathbf{r}) = - \frac{i}{2} \{ \nabla_{y}, \alpha(\mathbf{r})\} \sigma_{x} + \frac{i}{2} \{ \nabla_{x}, \alpha(\mathbf{r})\}\sigma_{y}.
 \end{equation}
Importantly,  $\hat{V}^{\mathrm{SOC}}(\mathbf{r})$ is a source of spin-momentum locking inhomogeneities, and plays an essential role in the bilinear system response. The parameter $\alpha(\mathbf{r})$ denotes spin-orbit coupling  related to local fluctuations of the spin-momentum locking
in the system. We assume a white noise distribution of these fluctuations, thus ${\langle \alpha(\mathbf{r}) \rangle = 0}$, ${\langle \alpha(\mathbf{r}) \alpha(\mathbf{r}')\rangle = n_{s} \alpha^{2} \delta(\mathbf{r} - \mathbf{r}')}$, and $\langle |\alpha_{\mathbf{k}\mathbf{k}'}|^{2} \rangle = n_{s} \alpha^{2}$.
\\
 The second common source of scattering processes are randomly distributed nonmagnetic and/or magnetic impurities, $\hat{V}^{\mathrm{IM}}$ \cite{Wang2007,Inoue2006,Nunner2008,Trushin2009,Wang2010,Bauer2017,Raimondi2019}. We assume that magnetic moments of magnetic impurities are aligned  with the external magnetic field. Thus, one may write:
 \begin{equation}
 V^{IM}(\mathbf{r}) = \left(V_{0}\, \sigma_{0} + V_{M}\,  \bsig \cdot \hat{\bver} \right) \sum_{l} \delta(\mathbf{r} - \mathbf{R}_{l}),
 \end{equation}
 where $\hat{\bver} = \mathbf{B}/B$ is a unit vector along the external magnetic field,  ${V_{M} = J s}$ describes the strength of exchange interaction between electrons and local spins $s$ ($J$ is an exchange constant), and $V_{0}$ describes the strength of scalar component of scattering potential. The magnetic moments of impurities contribute to the exchange potential, ${\langle V_{M}(\mathbf{r})\rangle = n_{i} V_{M} = M}$, where $n_i$ is the impurity concentration. The corresponding exchange  field $\mathbf{M}$ is parallel to the external magnetic field, and effectively leads to renormalization of this field, ${\mathbf{B} \rightarrow \tilde{\mathbf{B}} = \mathbf{B} + \mathbf{M} = (B + M) (\cos\phi_{B}, \sin \phi_{B})}$ and $\tilde{B} = B + M= g \mu_{B} \tilde{\mathcal{B}}$  ($ \tilde{\mathcal{B}}$ is given in Tesla). In addition, we assume the white noise distribution for the scalar component of scattering potential, i.e., ${\langle V_{0}(\mathbf{r})\rangle = 0}$, $\langle V_{0}(\mathbf{r}) V_{0}(\mathbf{r}') \rangle \neq 0$, and $\langle |V_{0\,\mathbf{k}\mathbf{k}'}|^2\rangle = n_{i} v_{0}^{2}$.
 \\
 The scattering potential defined in the momentum space takes the following form:
\begin{equation}
V^{\mathrm{scatt}}_{\mathbf{k} \mathbf{k}'} = V^{SOC}_{\mathbf{k} \mathbf{k}'} + V^{IM}_{\mathbf{k} \mathbf{k}'},
\end{equation}
where
\begin{equation}
V_{\mathbf{k}\mathbf{k}'}^{SOC} = \frac{\alpha}{2} \left[ (k_{y} + k'_{y}) \sigma_{x} - (k_{x} + k'_{x}) \sigma_{y}\right],
\end{equation}
and
\begin{equation}
\label{UMFT}
V^{IM}_{\mathbf{k} \mathbf{k}'} = v_{0} ( \sigma_{0} + \gamma_{x}\sigma_{x} + \gamma_{y} \sigma_{y}),
\end{equation}
with $\gamma_{x,y}$ describing the ratio of magnetic and scalar potentials, i.e., ${\gamma_{x} = \gamma \cos{\phi_{B}}}$, ${\gamma_{y} = \gamma \sin{\phi_{B}}}$, and $\gamma = V_{M}/v_{0}$.
\\

The in plane total effective magnetic field, $\tilde{\mathbf{B}}_{\mathrm{eff}} = \tilde{\mathbf{B}} + \mathcal{J} \mathbf{S}^{\mathbf{(E)}}$, can be simply removed from the Hamiltonian $\hat{H}_{0}$ by the gauge transformation \cite{Dyrdal_PRL2020,Bauer2017}: ${\mathbf{k} \rightarrow \mathbf{q} - \frac{e}{\hbar} \mathbf{\Lambda}}$, where $\mathbf{\Lambda} = \frac{\hbar}{\upsilon e} \hat{\mathbf{z}} \times \tilde{\mathbf{B}}_{\mathrm{eff}}$. However, after such a gauge transformation, the spin-orbit scattering potential takes the form which is $\tilde{\mathbf{B}}_{\mathrm{eff}}$--dependent. Thus, upon the gauge transformation the system Hamiltonian takes the form: ${\hat{H}_{\mathbf{q}\mathbf{q}'} = \hat{H}^{0}_{\mathbf{q}\mathbf{q}'} + \hat{V}^{\mathrm{scatt.}}_{\mathbf{q}\mathbf{q}'}} $, where $\hat{H}^{0}_{\mathbf{q}\mathbf{q}'} = \upsilon(\mathbf{q}\times\hat{\mathbf{z}})\cdot \mathbf{\bsig}\delta_{\mathbf q, \mathbf q'}$ and ${V^{\mathrm{scatt}}_{\mathbf{q} \mathbf{q}'} = V^{SOC}_{\mathbf{q} \mathbf{q}'} + V^{IM}_{\mathbf{q} \mathbf{q}'}}$, with $V^{IM}_{\mathbf{q} \mathbf{q}'}$ having the same form as before the transformation, and $V^{SOC}_{\mathbf{q} \mathbf{q}'}$ given by the formula:
\begin{eqnarray}
V^{SOC}_{\mathbf{q} \mathbf{q}'} =\frac{\alpha}{2} \left[ (q_{y} + q'_{y}) \sigma_{x} - (q_{x} + q'_{x}) \sigma_{y}\right]
\nonumber\\ - \frac{\alpha}{2} \left[ \tilde{B}_{x} \sigma_{x} - (\tilde{B}_{y} + \mathcal{J} S^{(E)}_{y}) \sigma_{y} \right].
\end{eqnarray}

\section{Boltzmann formalism in the case of anisotropic scattering}
In the problem under consideration, electron scattering is anisotropic and a simple relaxation time approximation is not adequate \cite{Schliemann2003,Vyborny2009}. Therefore, we use  the Boltzmann theory in its  general form and describe below solution of the generalized Boltzmann equation.
Accordingly, to obtain the longitudinal resistance and MR  we use the Boltzmann equation in the following well-known  form:
\begin{equation}
\label{eq:Boltzmann}
 e \mathbf{E} \cdot \mathbf{v}(\mathbf{q})\partial_{\varepsilon} f^{0}_{\varepsilon}(\mathbf{q}) = \int \frac{d^{2} \mathbf{q}'}{(2\pi )^{2}} w_{\mathbf{q}\mathbf{q}'} \left[ f_{\varepsilon}(\mathbf{q}, \mathbf{E}) - f_{\varepsilon}(\mathbf{q}', \mathbf{E})\right],
\end{equation}
where $e$ is the electron charge, $\mathbf{E}$ stands for electric field, ${\mathbf{v} = \frac{1}{\hbar} \nabla_{\mathbf{q}} \varepsilon_{q}}$ is the velocity for the band with the dispersion relation  $\varepsilon_{q}$, while $f^{0}_{\varepsilon}(\mathbf{q}) = f^{0}$ and $f_{\varepsilon}(\mathbf{q}, \mathbf{E}) = f$ define the equilibrium and  non-equilibrium distribution functions, respectively.
In general, the distribution function  depends on $\mathbf{q} = q (\cos\phi, \sin\phi)$ and  electric field $\mathbf{E}$. Assuming electric filed along the axis $x$,  $\mathbf{E}= (E, 0)$, and expanding the distribution function with respect to $E$, one can write in  the linear response:
\begin{equation}
\label{eq:f}
f= f^{0} + E \partial_{E} f .
\end{equation}
We rewrite this equation in the form
\begin{equation}
\label{eq:f-f0}
f - f^{0} = E  A(\phi),
\end{equation}
where the function $A(\phi)$ is defined as follows:
\begin{eqnarray}
A(\phi) = \partial_{E} f = e\, v\, a(\phi) \,\partial_{\varepsilon} f^{0}.
\end{eqnarray}
Then, combining Eq.(\ref{eq:f-f0}) with Eq.(\ref{eq:Boltzmann}) one gets:
\begin{equation}
\label{eq:cosphi}
\cos(\phi) = \bar{w}_\phi a(\phi) - \int d\phi' w_{\phi, \phi'}  a(\phi'),
\end{equation}
where
$\bar{w}_\phi = \int d \phi' w_{\phi, \phi'}$  and  $w_{\phi, \phi'} = \frac{1}{(2\pi )^{2}} \int q' dq' w_{\mathbf{q},\mathbf{q}'}$.
From  the above Fredholm equation one can determine $a(\phi)$,
which when inserted to Eq.(\ref{eq:f-f0}) allows to find the solution of the Boltzmann equation in the linear approximation with respect to electric field.
To do this we write Eq.(\ref{eq:cosphi}) as follows:
\begin{eqnarray}
\label{eq:a}
a(\phi) = \alpha(\phi) + \int d\phi' \mathcal{K}(\phi, \phi') a(\phi'),
\end{eqnarray}
where $\alpha(\phi) = \frac{\cos\phi}{\bar{w}_{\phi}}$ and $\mathcal{K}(\phi, \phi') = \frac{w_{\phi,\phi'}}{\bar{w}_{\phi}}$.
The scattering rate $w_{\phi, \phi'} $ and $\mathcal{K}(\phi, \phi') $ can be cast as:
\begin{eqnarray}
w_{\phi, \phi'} =  \frac{q}{2 h \upsilon } \sum_{j}^{3} \lambda_{j}(\phi) \kappa_{j}(\phi'),\\
\label{eq:Kphiphip}
\mathcal{K}(\phi, \phi')  = \sum_{j}^{3} \frac{\lambda_{j}(\phi)}{2 \pi \lambda_{3}(\phi)} \kappa_{j}(\phi'),
\end{eqnarray}
where the functions $\kappa_{j}(\phi')$ for $j = 1,2,3$ have the following form:
\begin{eqnarray}
\kappa_{1}(\phi') = \cos\phi',\quad \kappa_{2}(\phi') = \sin\phi', \quad \kappa_{3}(\phi') = 1,
\end{eqnarray}
The functions  $\lambda_{j}$ appearing in the kernel of integral equation (\ref{eq:a}) have the following form,
\begin{eqnarray}
\lambda_1(\phi)=\mu_3\cos\phi+\mu_4\sin\phi+\mu_2,\\
\lambda_2(\phi)=\mu_4\cos\phi+\mu_5\sin\phi+\mu_1,\\
\lambda_3(\phi)=\mu_2\cos\phi+\mu_1\sin\phi+\mu_0,
\label{lambda}
\end{eqnarray}
where,
{\small{
\begin{align}
\label{mu}
\mu_5 &=n_{S}\alpha^2 q^2\left(1-\left(\frac{\mathcal{J}S^{(E)}_y}{\upsilon q}\right)^2+\left( \frac{\tilde{B}}{\upsilon q} \right)^2\cos2\phi_B\right.\nonumber\\
& \hspace{0.8cm}-\left.2 \frac{\mathcal{J}S^{(E)}_{y} \tilde{B}}{\upsilon^{2} q^{2}} \sin\phi_B\right)+ n_{i}v_0^2\left(1+\gamma^2\cos2\phi_B\right),\\
\mu_4 &=- n_{S}\alpha^2 q^2\left(\left( \frac{\tilde{B}}{\upsilon q} \right)^2\sin 2\phi_B + 2 \frac{\mathcal{J}S^{(E)}_{y} \tilde{B}}{\upsilon^{2} q^{2}}\cos\phi_B\right) \nonumber\\ 
& \hspace{4.5cm}-n_{i}v_0^2\gamma^2\sin2\phi_B,\\
\mu_3 &=n_{S}\alpha^2 q^2\left(1+\left(\frac{\mathcal{J}S^{(E)}_y}{\upsilon q}\right)^2-\left( \frac{\tilde{B}}{\upsilon q} \right)^2\cos2\phi_B\right. \nonumber\\
&\hspace{0.8cm}+\left.2 \frac{\mathcal{J}S^{(E)}_{y} \tilde{B}}{\upsilon^{2} q^{2}}\sin\phi_B\right)+ n_{i}v_0^2\left(1-\gamma^2\cos2\phi_B\right),\\
\mu_2 &=-n_{S}\alpha^2 q^2\left(2\frac{\mathcal{J}S^{(E)}_y}{\upsilon q}+2 \frac{\tilde{B}}{\upsilon q}\sin\phi_B\right)-2n_{i}v_0^2\gamma\sin\phi_B,\hspace{0.5cm}\\
\mu_1 &=2n_{S}\alpha^2q^2  \frac{\tilde{B}}{\upsilon q}\cos\phi_B+2n_{i}v_0^2 \gamma\cos\phi_B,\hspace{0.3cm}\\
\mu_0 &=n_{S}\alpha^2 q^2\left(1+\left(\frac{\mathcal{J}S^{(E)}_y}{\upsilon q}\right)^2+\left( \frac{\tilde{B}}{\upsilon q} \right)^2\right. \nonumber\\
&\hspace{1.5cm}+\left.2 \frac{\mathcal{J}S^{(E)}_{y} \tilde{B}}{\upsilon^{2} q^{2}}\sin\phi_B\right)+n_{i}v_0^2\left(1+\gamma^2\right).
\end{align}
}}

Combining  (\ref{eq:Kphiphip}) with (\ref{eq:a}) one finds
\begin{align}
a(\phi) &= \alpha(\phi) + \sum_{j}^{3} \frac{\lambda_{j}(\phi)}{2\pi \lambda_{3}(\phi)} c_{j}, \\
c_{j} &= \int d\phi' \kappa_{j}(\phi') a(\phi').
\end{align}
Then, upon multiplying this equation by $\kappa_{i}(\phi)$ and integrating over $\phi$ one obtains
\begin{equation}
c_{i} =  d_{i} + \sum_{j}^{3} c_{j} m_{ji},
\end{equation}
where $d_{i} = \int d\phi\, \alpha(\phi) \kappa_{i}$ and $m_{ji} = \int d\phi\, \frac{\lambda_{j}(\phi)}{\lambda_{3}(\phi)} \kappa_{i}$. Thus the Fredholm equation can be rewritten as a matrix equation:
\begin{equation}
(\check{M} - \check{1}) \check{C} + \check{D} = 0,
\end{equation}
 where $m_{ij}$ are the matrix elements of $\check{M}$, whereas $ \check{C} $ and $\check{D}$ are vectors formed by $c_{i}$ and $d_{i}$, respectively.
Determining the coefficients $c_{j}$, we obtain  an explicit form of  $a(\phi)$, and thus of the distribution function.

The charge current density is given by the following formula:
\begin{equation}
\mathbf{j} = e \int \frac{d^2 \mathbf{q}}{(2\pi )^{2}} \mathbf{v}(\mathbf{q}) f(\mathbf{q}, \mathbf{E}).
\end{equation}
Thus, according to the notation introduced above, the longitudinal charge conductivity can be written as
$\sigma_{xx} = \frac{1}{E}j$,
which leads to the following expression:
\begin{equation}
\sigma_{xx} = e^{2} \int \frac{dq d\phi}{(2\pi)^{2}} v^{2}(q)\,q\, (\partial_{\varepsilon} f^{0})\, a(\phi)\, \cos\phi.
\end{equation}
In low temperature limit the above expression takes simple form:
\begin{equation}
\sigma_{xx} = \frac{e^{2}}{h} v_{F} q_{F} \int \frac{d\phi}{2\pi} a(q_{F}, \phi)\cos\phi ,
\end{equation}
where $q_F$ is the Fermi wavevector. 
This expression is a basis for numerical calculations presented and discussed in the next section.

\section{Bilinear magnetoresistance}
\begin{figure}[t]
\centerline{\includegraphics[width=0.85\columnwidth]{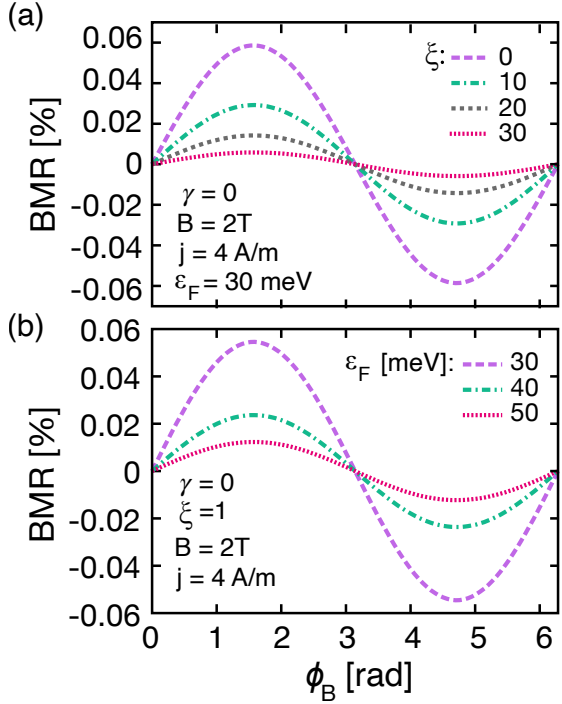}}
\caption{ BMR in the absence of magnetic impurities, $\gamma = 0$, as a function of the angle defining orientation of external magnetic field, $\phi_{B}$,  for  (a) selected values of the strength of scalar potential, $\xi = n_i v_{0}^{2}/[n_i v_{0}^{2}]_{\mathrm{ref}}$ for the Fermi energy $\varepsilon_{F}=30$ meV, and for (b) selected values of Fermi energies  and  $\xi = 1$. The other parameters: $n_{s} \alpha^{2} = 81\times10^{-40} {\mathrm{eV^{2} m^{4}}}$, $\tilde{\mathcal{B}}=2$T and $j=4$A/m.}\label{fig:bmr_qmr_x}
\end{figure}

The magnetoresistance is defined by the simple formula:
\begin{equation}
MR = \frac{\rho_{xx}(B) - \rho_{xx}(B=0)}{\rho_{xx}(B=0)} = \frac{\sigma_{xx}(B=0)}{\sigma_{xx}(B)} -1.
\end{equation}
Due to the nonequilibrium spin polarization, which is proportional to the charge current density, i.e., $S_{y} = \frac{\hbar^{2}}{2 ev} j_{x}$, the longitudinal conductivity depends also on $j_{x}$. Accordingly, one can define the antisymmetric (linear in both magnetic field and charge current density) magnetoresistance response, that is called the bilinear magnetoresistance, BMR:
\begin{eqnarray}
BMR = \frac{1}{2}  \left[ MR(j_{x} = j) - MR(j_{x} = -j) \right]
\end{eqnarray}

Using Eq. (24) we obtained BMR numerically.
Figure 1 presents  BMR  for the system in the absence of magnetic scattering on impurities, i.e., for $\gamma = 0$. The parameter $\xi$  is defined as $\xi = n_i v_{0}^{2}/[n_i v_{0}^{2}]_{\mathrm{ref}}$ where $[n_i v_{0}^{2}]_{\mathrm{ref}}$ is the reference value, i.e., $[n_i v_{0}^{2}]_{\mathrm{ref}} = 1.58 \times 10^{-24} {\mathrm{eV^{2}m^{2}}}$. Thus,  $\xi$ defines the strength of scalar impurity potential. BMR oscillates with the angle $\phi_{B}$ with the oscillation period equal to $2\pi$. This figure clearly shows the unidirectional behaviour of BMR. In addition, the amplitude of BMR decreases with increasing scattering on the scalar impurity potential and with increasing Fermi energy, as  reported previously.
Figure 2 shows BMR in the presence of scattering on magnetic impurities, $\gamma > 0$. As long as the magnetic part of the scattering potential is small in comparison to the scalar term, $\gamma < 1$, there are no qualitative changes in the behavior of BMR. 
\begin{figure}[t]
\centerline{\includegraphics[width=0.85\columnwidth]{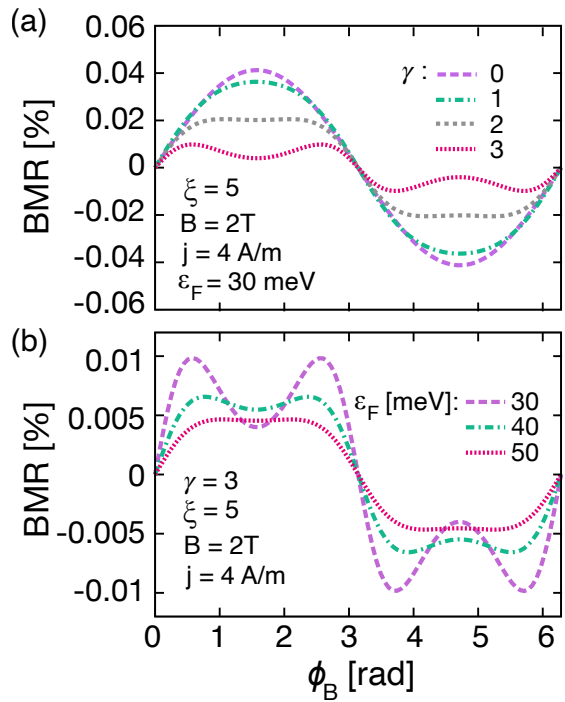}}
\caption{ BMR, in the presence of magnetic impurities, as a function of the angle defining orientation of external magnetic field, $\phi_{B}$, for (a)  selected  values of the magnetic impurity potential ($\gamma>0$) and fixed strength of the scalar potential, $\xi= 5$, and for (b) selected values of Fermi energy and fixed scalar and magnetic impurity potentials. The other parameters are the same as in Fig.1. The curve for $\gamma = 0$ is added to highlight the role of magnetic impurities.
} 
\label{fig:bmr_qmr_x}
\end{figure}
When the magnetic part of impurity potential becomes dominant, $\gamma > 1$, there are qualitative changes in the bilinear magnetoresistance. BMR oscillates then with the period $2\pi$, but for $\gamma >2$ one can identify four peaks that appear for $\phi_{B} \approx (2n +1) \pi/4$. Thus BMR can be expressed as $BMR \sim \mathcal{F}_{S} \sin {\phi_{B}} + \mathcal{F}_{M} \sin {\phi_{B}}\cos^2{\phi_{B}}$.
 To account for this behavior, we note that scattering on magnetic impurities in the absence of spin-orbit scattering cannot generate BMR, though magnetic part of the scattering potential can lead to  anisotropic magnetoresistance.  Accordingly, magnetic scattering ($\gamma >0$) on magnetic impurities must lead to suppression of BMR, similarly as scattering on the scalar potential. However, due to the anisotropic character of the this scattering,  the BMR effectively acquires additional oscillation period $\pi$.

The presented results show that BMR can be tuned by the concentration of magnetic impurities. Thus, apart from using BMR  in the new generation of logic devices, it can also be used as a tool for the characterization of transport properties in systems with strong spin-orbit coupling.

\section{Conclusions}

We have considered the bilinear magnetoresistance of  surface electron states in 3D topological insulators. To describe the topological electron states we have used the minimal model of surface states in TIs, that assumes isotropic Fermi contour and the absence of hexagonal warping. To describe scattering process we assumed spin-momentum locking inhomogeneities and also randomly distributed magnetic impurities with the magnetic moments oriented along the external magnetic field.  We have obtained the unidirectional character of MR that is a consequence of non-equilibrium spin polarization, spin-momentum locking inhomogeneity and scattering on magnetic impurities. We have also shown that the interplay of scattering on magnetic impurities and on spin-momentum locking inhomogeneities leads to a significant modification of the unidirectional magnetoresistance signal in comparison to that in the absence of magnetic impurities, especially for a significant  magnetic part of impurity scattering potential.  A new feature of BMR is then an additional period in the oscillation of BMR with the angle between magnetic field and current.

\section*{Acknowledgments}
We thank Prof. Józef Barnaś for reading this manuscript and valuable discussions.
This work has been supported by the Norwegian Financial Mechanism under the Polish-Norwegian Research Project NCN GRIEG “2Dtronics”, project no. 2019/34/H/ST3/00515.
%



  \bibliography{Ref.bib}

\end{document}